\documentclass[10pt,letterpaper]{article}
\usepackage{opex3}
\usepackage{epsfig}
\usepackage{amsmath}
\usepackage{amssymb}

\begin{document}

\title{Predicting macrobending loss for large-mode area photonic crystal fibers}

\author{M.~D. Nielsen,$^{1,2,*}$ N.~A. Mortensen,$^{1,3}$ M. Albertsen,$^{2,4}$ \\ J.~R. Folkenberg,$^1$ A. Bjarklev,$^2$ and D. Bonacinni$^4$}
\address{$^1$Crystal Fibre A/S, Blokken 84, DK-3460 Birker\o d, Denmark\\$^2$COM, Technical University of Denmark, DK-2800 Kongens Lyngby, Denmark\\$^3$Department of Micro and Nanotechnology, Technical University of Denmark,\\ DK-2800 Kongens Lyngby, Denmark\\$^4$European Southern Observatory, Karl-Swarzschildstrasse 2,\\ D-85748 Garching bei M\"{u}nchen, Germany}
\email{$^*$mdn@crystal-fibre.com}


\begin{abstract}
We report on an easy-to-evaluate expression for the prediction of
the bend-loss for a large mode area photonic crystal fiber (PCF)
with a triangular air-hole lattice. The expression is based on a
recently proposed formulation of the V-parameter for a PCF and
contains no free parameters.  The validity of the expression is
verified experimentally for varying fiber parameters as well as
bend radius. The typical deviation between the position of the
measured and the predicted bend loss edge is within measurement
uncertainty.
\end{abstract}




\section{Introduction}
\begin{figure}[b!]
\begin{center}
\epsfig{file=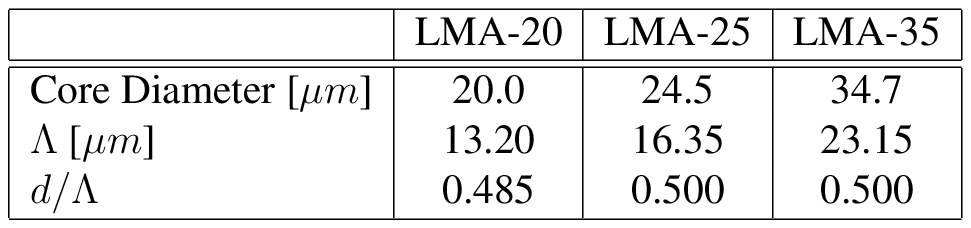,
height=0.17\textwidth}\hspace{5mm}\epsfig{file=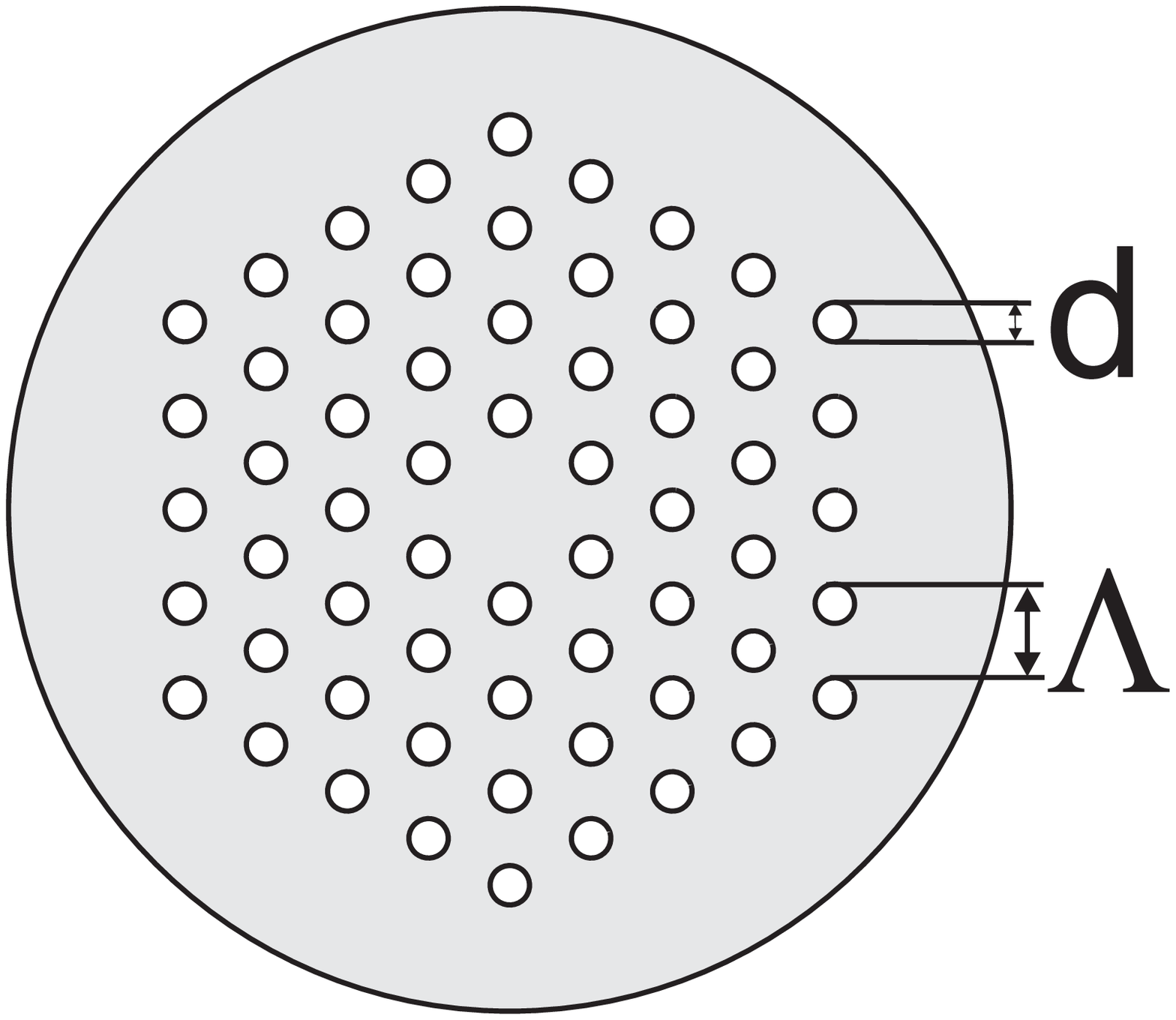,
height=0.17\textwidth}
\end{center}
\caption{Structural data for the LMA fibers which all have a cross-section with a triangular arrangement of air-holes running along the full length of the fiber.}
\label{fig1}
\end{figure}

In solid-core photonic crystal fibers (PCF) the air-silica
microstructured cladding (see Fig.~\ref{fig1}) gives rise to a
variety of novel phenomena~\cite{knight2003} including large-mode
area (LMA) endlessly-single mode operation \cite{birks1997}.
Though PCFs typically have optical properties very different from
that of standard fibers they of course share some of the overall
properties such as the susceptibility of the attenuation to
macro-bending.

Macrobending-induced attenuation in PCFs has been addressed both
experimentally as well as theoretically/numerically in a number of
papers
\cite{birks1997,sorensen2001,sorensen2002,mortensen2003b,baggett2003}.
However, predicting bending-loss is no simple task and typically
involves a full numerical solution of Maxwell's equations as well
as use of a phenomenological free parameter, {\it e.g.} an
effective core radius. In this paper we revisit the problem and
show how macro-bending loss measurements on high-quality PCFs can
be predicted with high accuracy using easy-to-evaluate empirical
relations.

\section{Predicting macro-bending loss}

Predictions of macro-bending induced attenuation in photonic
crystal fibers have been made using various approaches including
antenna-theory for bent standard fibers
\cite{sorensen2001,sorensen2002}, coupling-length criteria
\cite{birks1997,mortensen2003b}, and phenomenological models
within the tilted-index representation \cite{baggett2003}. Here,
we also apply the antenna-theory of Sakai and
Kimura~\cite{sakai1978,sakai1979}, but contrary to
Refs.~\cite{sorensen2001,sorensen2002} we make a full
transformation of standard-fiber parameters such as $\Delta$, $W$,
and $V$ \cite{snyder} to fiber parameters appropriate to
high-index contrast PCFs with a triangular arrangement of air
holes. In the large-mode area limit we get (see Appendix)

\begin{equation}\label{alpha_LMA}
\alpha\Lambda\simeq \frac{1}{8 \sqrt{6\pi}}\frac{1}{
n_S}\frac{\Lambda^2}{A_{\rm eff}}\frac{\lambda}{\Lambda} {F}\left(
\frac{1}{6\pi^2}\frac{1}{
n_S^2}\frac{R}{\Lambda}\left(\frac{\lambda}{\Lambda}\right)^2
V^3_{\rm PCF}\right)\;,\; {F}(x)=x^{-1/2}\exp(-x),
\end{equation}
for the power-decay, $P(z)=P(0)\exp(-2\alpha z)$, along the fiber.
For a conversion to a dB-scale $\alpha$ should be multiplied by
$20\times\log_{10}(e)\simeq 8.686$. In Eq.~(\ref{alpha_LMA}), $R$
is the bending radius, $A_{\rm eff}$ is the effective area
\cite{mortensen2002a}, $n_S$ is the index of silica, and
\begin{equation}\label{V_PCF}
V_{\rm PCF}=\Lambda\sqrt{\beta^2-\beta_{cl}^2}
\end{equation}
is the recently introduced effective V-parameter of a
PCF~\cite{mortensen2003c}. The strength of our formulation is that
it contains no free parameters (such as an arbitrary core radius)
and furthermore empirical expressions, depending only on
$\lambda/\Lambda$ and $d/\Lambda$, have been given recently for
both $A_{\rm eff}$ and $V_{\rm PCF}$
\cite{nielsen2003b,nielsen2003c}.

From the function ${F}(x)$ we may derive the parametric dependence
of the critical bending radius $R^*$. The function increases
dramatically when the argument is less than unity and thus we may
define a critical bending radius from $x\sim 1$ where ${F}\sim
1/e$. Typically the PCF is operated close to cut-off where $V_{\rm
PCF}^*=\pi$~\cite{mortensen2003c} so that the argument may be
written as

\begin{equation}
\underbrace{\pi^3\frac{1}{6\pi^2}\frac{1}{ n_S^2}}_{\sim 1/4}\frac{R^*}{\Lambda}\left(\frac{\lambda}{\Lambda}\right)^2
\sim 1 \Rightarrow R^*\propto \frac{\Lambda^3}{ \lambda^2}
\end{equation}
This dependence was first reported and experimentally confirmed by Birks {\it et al.}~\cite{birks1997} and recently a pre-factor of order unity was also found experimentally in Ref.~\cite{mortensen2003b}.

\begin{figure}[t!]
\begin{center}
\epsfig{file=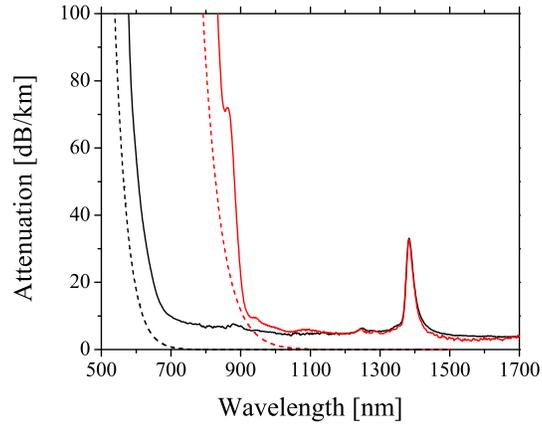, width=0.6\textwidth,clip}
\end{center}
\caption{Macro-bending loss for the LMA-20 fiber for bending radii
of R=8 cm (red, solid curve) and R=16 cm (black, solid curve).
Predictions of Eq.~(\ref{alpha_LMA}) are also included (dashed
curves).} \label{fig2}
\end{figure}

\begin{figure}[b!]
\begin{center}
\epsfig{file=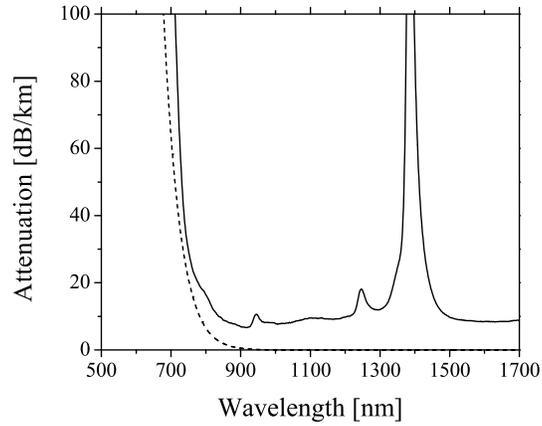, width=0.6\textwidth,clip}
\end{center}
\caption{Macro-bending loss for the LMA-25 fiber for bending
radius of R=16 cm (solid curve). Predictions of
Eq.~(\ref{alpha_LMA}) are also included (dashed curve).}
\label{fig3}
\end{figure}

\section{Experimental results}

We have fabricated three LMA fibers by the stack-and-pull method
and characterized them using the conventional cut-back technique.
All three fibers have a triangular air-hole array and a solid core
formed by a single missing air-hole in the center of the
structure, see Fig.~\ref{fig1}.

\begin{figure}[t!]
\begin{center}
\epsfig{file=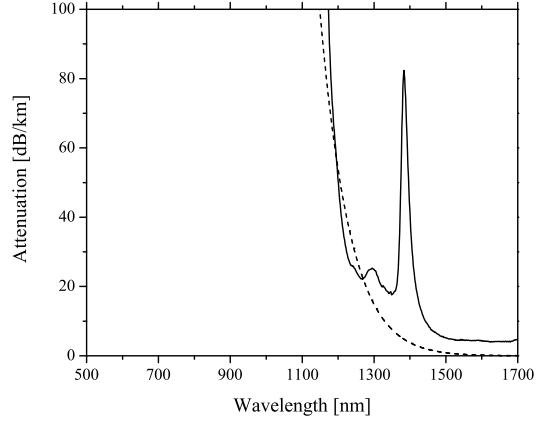, width=0.6\textwidth,clip}
\end{center}
\caption{Macro-bending loss for the LMA-35 fiber for bending
radius of R=16 cm (solid curve). Predictions of
Eq.~(\ref{alpha_LMA}) are also included (dashed curve).}
\label{fig4}
\end{figure}

For the LMA-20 macro-bending loss has been measured for bending
radii of R=8 cm and R=16 cm and the results are shown in
Fig.~\ref{fig2}. The predictions of Eq.~(\ref{alpha_LMA}) are also
included. It is emphasized that the predictions are based on the
empirical relations for $A_{\rm eff}$ and $V_{\rm PCF}$ provided
in Refs. \cite{nielsen2003b} and \cite{nielsen2003c} respectively
and therefore do not require any numerical calculations. Similar
results are shown in Figs.~\ref{fig3} and \ref{fig4} for the
LMA-25 and LMA-35 fibers, respectively.

\section{Discussion and conclusion}

The PCF, in theory, exhibits both a short and long-wavelength
bend-edge. However, the results presented here only indicate a
short-wavelength bend-edge. The reason for this is that the
long-wavelength bend-edge occurs for $\lambda \gg \Lambda/2$
\cite{sorensen2001}. For typical LMA-PCFs it is therefor located
in the non-transparent wavelength regime of silica.

 In conclusion we have demonstrated that
macro-bending loss measurements on high-quality PCFs can be
predicted with good accuracy using easy-to-evaluate empirical
relations with only $d$ and $\Lambda$ as input parameters. Since
macro-bending attenuation for many purposes and applications is
the limiting factor we believe that the present results will be
useful in practical designs of optical systems employing photonic
crystal fibers.

\section*{Appendix}

 The starting point is the bending-loss formula for a Gaussian mode in a standard-fiber \cite{sakai1978,sakai1979}

\begin{equation}\label{alpha1}
\alpha=\frac{\sqrt{\pi}}{8}\frac{1}{A_{\rm eff}}\frac{\rho}{W}\frac{\exp\left(-\frac{4}{3}\frac{R}{\rho}\frac{\Delta}{V^2}W^3\right)}{\sqrt{W\frac{R}{\rho}+\frac{V^2}{2\Delta W}}}
\end{equation}
where $A_{\rm eff}$ is the effective area, $\rho$ is the core radius, $R$ is the bending radius, and the standard-fiber parameters are given by \cite{sakai1978,snyder}
\begin{equation}\label{standard-fib}
\Delta=\frac{sin^2\theta_c}{2}\;,\;V=\beta\:\rho\:sin\theta_c\;,\;W=\rho\sqrt{\beta^2-\beta_{cl}^2}.
\end{equation}
Substituting these parameters into Eq.~(\ref{alpha1}) we get

\begin{equation}\label{alpha2}
\alpha\Lambda\simeq
\frac{1}{8}\sqrt{\frac{2\pi}{3}}\frac{\Lambda^2}{A_{\rm
eff}}\frac{1}{\beta\Lambda}{F}\left(\frac{2}{3}\frac{R}{\Lambda}\frac{V^3_{\rm
PCF}}{\left(\beta\Lambda\right)^2}\right)
\end{equation}
in the relevant limit where $R\gg \rho$. Here, $F$ and $V_{\rm
PCF}$ in Eqs.~(\ref{alpha_LMA}) and (\ref{V_PCF}) have been
introduced. For large-mode area fibers we make a further
simplification for the isolated propagation constant; using that
$\beta = 2\pi n_{\rm eff}/\lambda \simeq 2\pi n_S /\lambda$ we
arrive at Eq.~(\ref{alpha_LMA}).

\section*{Acknowledgments}

M.~D. Nielsen acknowledges financial support by the Danish Academy
of Technical Sciences.


\begin{thebibliography}{10}
\newcommand{\enquote}[1]{``#1''}
\expandafter\ifx\csname url\endcsname\relax
  \def\url#1{\texttt{#1}}\fi
\expandafter\ifx\csname
urlprefix\endcsname\relax\def\urlprefix{URL }\fi
\providecommand{\eprint}[2][]{\url{#2}}

\bibitem{knight2003}
J.~C. Knight, \enquote{Photonic crystal fibres,} Nature
\textbf{424}, 847--851
  (2003).

\bibitem{birks1997}
T.~A. Birks, J.~C. Knight, and P.~S.~J. Russell,
\enquote{Endlessly single mode
  photonic crystal fibre,} Opt. Lett. \textbf{22}, 961--963 (1997).

\bibitem{sorensen2001}
T.~S{\o}rensen, J.~Broeng, A.~Bjarklev, E.~Knudsen, and S.~E.~B.
Libori,
  \enquote{Macro-bending loss properties of photonic crystal fibre,} Electron.
  Lett. \textbf{37}, 287--289 (2001).

\bibitem{sorensen2002}
T.~S{\o}rensen, J.~Broeng, A.~Bjarklev, T.~P. Hansen, E.~Knudsen,
S.~E.~B.
  Libori, H.~R. Simonsen, and J.~R. Jensen, \enquote{Spectral Macro-bending
  loss considerations for photonic crystal fibres,} IEE Proc.-Opt.
  \textbf{149}, 206 (2002).

\bibitem{mortensen2003b}
N.~A. Mortensen and J.~R. Folkenberg, \enquote{Low-loss criterion
and effective
  area considerations for photonic crystal fibers,} J. Opt. A: Pure Appl. Opt.
  \textbf{5}, 163--167 (2003).

\bibitem{baggett2003}
J.~C. Baggett, T.~M. Monro, K.~Furusawa, V.~Finazzi, and D.~J.
Richardson,
  \enquote{Understanding bending losses in holey optical fibers,} Opt. Commun.
  \textbf{227}, 317--335 (2003).

\bibitem{sakai1978}
J.~Sakai and T.~Kimura, \enquote{Bending loss of propagation modes
in
  arbitrary-index profile optical fibers,} Appl. Opt. \textbf{17}, 1499--1506
  (1978).

\bibitem{sakai1979}
J.~Sakai, \enquote{Simplified bending loss formula for single-mode
optical
  fibers,} Appl. Opt. \textbf{18}, 951--952 (1979).

\bibitem{snyder}
A.~W. Snyder and J.~D. Love, \emph{Optical Waveguide Theory}
(Chapman \& Hall,
  New York, 1983).

\bibitem{mortensen2002a}
N.~A. Mortensen, \enquote{Effective area of photonic crystal
fibers,} Opt.
  Express \textbf{10}, 341--348 (2002).
  \urlprefix\url{http://www.opticsexpress.org/abstract.cfm?URI=OPEX-10-7-341}.

\bibitem{mortensen2003c}
N.~A. Mortensen, J.~R. Folkenberg, M.~D. Nielsen, and K.~P.
Hansen,
  \enquote{Modal cut-off and the $V$--parameter in photonic crystal fibers,}
  Opt. Lett. \textbf{28}, 1879--1881 (2003).

\bibitem{nielsen2003b}
M.~D. Nielsen, N.~A. Mortensen, J.~R. Folkenberg, and A.~Bjarklev,
  \enquote{Mode-Field Radius of Photonic Crystal Fibers Expressed by the
  $V$--parameter,} Opt. Lett. \textbf{28}, 2309--2311 (2003).

\bibitem{nielsen2003c}
M.~D. Nielsen and N.~A. Mortensen, \enquote{Photonic crystal fiber
design based
  on the $V$--parameter,} Opt. Express \textbf{11}, 2762--2768 (2003).
  \urlprefix\url{http://www.opticsexpress.org/abstract.cfm?URI=OPEX-11-21-2762%
}.

\end{thebibliography}
\end{document}